\documentclass[varenna]{cimento}

\usepackage{amssymb,amsfonts,bm}
\usepackage{graphics,graphicx}
\usepackage{sidecap,verbatim}

\title{Like-charge colloidal attraction: a simple argument}

\author{E. Trizac}

\institute{Laboratoire de Physique Th\'eorique et Mod\`eles Statistiques, 
UMR CNRS 8626, Universit\'e Paris-Sud, 91405 Orsay, France}

\author{L. \v{S}amaj}

\institute{Institute of Physics, Slovak Academy of Sciences, D\'ubravsk\'a Cesta 9,
Bratislava 84511, Slovakia}

\shortauthor{L. \v{S}amaj \atque E. Trizac}

\begin{document}

\maketitle

\begin{abstract}
By a length scale analysis, we study the equilibrium 
interactions between
two like-charge planes confining neutralising counter-ions.
At large Coulombic couplings, approaching the two charged bodies 
leads to an unbinding of counter-ions, a situation that is amenable
to an exact treatment.
This phenomenon is the key to attractive
effective interactions.
A particular effort is made for pedagogy,
keeping equations and formalism to a minimum.
\end{abstract}

\section{Introduction}
\label{sec:intro}
Consider two identical charged macromolecules alone in an electrolyte,
and assume that Coulombic forces exclusively are at work. By integrating out
the microscopic degrees of freedom (the electrolyte), one obtains the effective 
pair potential between the two macromolecules \cite{Bell00}: is it attractive
or repulsive? Of course in vacuum (i.e. without the electrolyte), 
the two bodies always repel, but the presence of the electrolyte complicates
the matter, to such an extent that the answer to the above question has been
a rather controversial subject since at least the 1930s 
\cite{LeDu39,VeOv48,LeHa92,Over93}. 
In essence, all early contributions pertained to mean-field,
a treatment for which it can be rigorously shown that the
effective potential is repulsive \cite{Neu99,SaCh00,Triz00}. 
Yet, it has been understood since the 1980s that like-charge attraction 
is nevertheless possible \cite{GJWL84,KjMa84,KMSS93}. Since this phenomenon
is a signature of non mean-field effects, it is not straightforward to
build an intuitive picture of the underlying mechanisms. It is our goal here to
present a simple argument where the prerequisite for like-charge attraction
become clear, and which provides exact results in some limiting sense
to be precised below (namely for short distances and large couplings). 

The paper is organised as follows. We first lay out the model in section
\ref{sec:model} where the contact theorem is also reminded. It is the 
cornerstone of our analysis. We then
analyse the limiting case where the two
macromolecules are distant (section \ref{sec:larged}),
which sheds interesting light on the physics
at work when the two plates are close, as developed in
section \ref{sec:smalld}. Our results are further discussed in
section \ref{sec:discuss} and conclusions are finally drawn 
in section \ref{sec:concl}.

\begin{SCfigure}[1.2]
\centering
\includegraphics[width=0.25\textwidth]{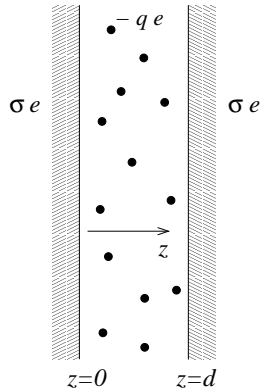}
\caption{Sketch of the geometry considered, and definition of relevant
parameters. 
The two charged parallel plates at distance $d$ are neutralised
by counter-ions of charge $-q e$, represented by the black dots.
The system can be viewed as a minimal model where 
like-charge attraction may be present, in some adequate range
of distance and coupling.}
\label{fig:geometry} 
\end{SCfigure}

\section{The model}
\label{sec:model}
We shall concentrate on the simplest setting possible, that where the two
macromolecules (colloids) are envisioned as two parallel planar surfaces  at 
a distance $d$, see fig.\ref{fig:geometry}. 
Each colloid/plate is assumed to bear a uniform surface 
charge distribution $\sigma e$, where $e$ is the elementary charge. The solvent
is a structure-less medium of dielectric permittivity $\epsilon$,
the same as the permittivity of the macromolecules themselves 
(dielectric image effects are discarded) and the system is in equilibrium at
temperature $T$. In between the colloids, mobile counter-ions of charge $-qe$ 
ensure global electro-neutrality.
While the microscopic density of counter-ions 
is strongly modulated (ions repel),
the coarse-grained density profile $\rho(z)$, averaged over a plane
at a given position $z$, is a smooth function. For the
sake of the argument, the counter-ions are assumed point-like: 
usually, a hard core is required to prevent collapse 
of oppositely charged small ions, but we deal here with a system 
having one type of small ions only, all of the same charge (no salt),
and that is therefore free of divergences.
From the permittivity $\epsilon$ and temperature $T$, we define the
so-called Bjerrum length $\ell_B = e^2 / (\epsilon kT)$ as the
distance where the repulsion between two elementary charges
coincides with thermal energy $kT$ ($k$ is Boltzmann's constant,
and we use CGS units, so that for water at room temperature,
one has $\ell_B \simeq 7 \,$\AA). Note that global electro-neutrality
imposes the following constraint on the counter-ion 
profile: $\int_0^d \rho(z)\,dz = 2\sigma/q$.

The question is: what is the sign of the pressure $P$, i.e. the force
between the two plates per unit surface ? To find the answer, it is essential 
here to invoke a key result, known as the contact theorem \cite{BlHL79}.
It states that $P$ is, quite intuitively, the sum of two contributions:
\begin{equation}
P \, = \, \rho(0) \, kT \, -\, \frac{2 \pi}{\epsilon} \,\sigma^2 e^2 
\label{eq:contact}
\end{equation}
The first one, $\rho(0)  kT$, is repulsive and stems from the collisions
of counter-ions with the plate. The second, 
$2 \pi \sigma^2 e^2/\epsilon $, is the traditional electrostatic
pressure, important in capacitor physics.
It is attractive, and simply means here that 
the plate located at $z=0$ and having surface charge $\sigma e$ 
sees on its right ($z>0$) an integrated charge
$\sigma e - qe \int_0^d \rho(z) dz =-\sigma e$ from electro-neutrality,
that is thus of opposite sign. The contact theorem (\ref{eq:contact})
is exact, and remains true when counter-ion excluded volume is accounted
for. We note that it implies a strong constraint on the contact density
$\rho(0)$ of an isolated macromolecule. Indeed, when $d\to \infty$,
the pressure $P$ should vanish so that eq. (\ref{eq:contact}) implies
$\rho(0) = 2 \pi \ell_B \sigma^2$. This 
result shows that the contact density does not depend on the valency $q$
of ions. When $d$ is finite, it is no longer possible to know 
$\rho(0)$ without an explicit and often complicated statistical mechanics
treatment, and the ensuing pressure is a non trivial quantity. 
It however only depends on 2 parameters (a reduced distance plus 
a coupling parameter)
and we shall see that for strongly correlated systems,
meaning for instance that $\sigma$ is large, and for small enough $d$,
$\rho(0)$ becomes simple again and yields an interesting equation 
of state through (\ref{eq:contact}). Before addressing that 
small $d$ situation, it is informative to
discuss in more details the case where the two plates are at a large 
distance from each other.

\section{The large distance limit}
\label{sec:larged}
When $d$ is large (meaning $d \gg a$ where $a$ is the lateral correlation
length which we are about to define), 
the ionic distribution $\rho(z)$
around a macromolecule is only weakly affected by the presence of the
other plate. We have furthermore emphasised in the introduction that the
relevant situation for like-charge attraction is that where
non mean-field effects are at work, so that the electrostatic
coupling should be large enough. To quantify such a coupling,
one can assume that all ions are collapsed onto the plate 
(which happens at low enough temperature), in which case the mean surface
per counter-ion is $q/\sigma$, which defines a typical distance 
$a\propto\sqrt{q/\sigma}$ between ions. Comparing their typical Coulombic repulsion 
$q^2 e^2/(\epsilon a)$ to $kT$, we obtain the ratio
\begin{equation}
\frac{q^2 e^2}{\epsilon a} \, \frac{1}{kT} \, \propto \, \frac{ e^2}{\epsilon kT} \,
q^2 \,\frac{\sigma^{1/2}}{q^{1/2}} \,=\, (q^3 \ell_B^2 \sigma)^{1/2}.
\end{equation}
Taking the square, 
we define the coupling constant $\Xi=2 \pi q^3 \ell_B^2 \sigma$
\cite{Levi02,NKNP10}.
This parameter discriminates weakly coupled cases where 
$\Xi \ll 1$ and mean-field holds, from strongly coupled ones,
where $\Xi \gg 1$ \cite{Levi02,Mess09,NKNP10}. 
Large couplings may be obtained in practice by increasing  $\sigma$
or $q$ (considering multivalent ions such as Cr$^{3+}$,
spermidine$^{3+}$ having $q=3$, or spermine$^{4+}$ with $q=4$.)

\begin{SCfigure}
\centering
\includegraphics[width=0.59\textwidth]{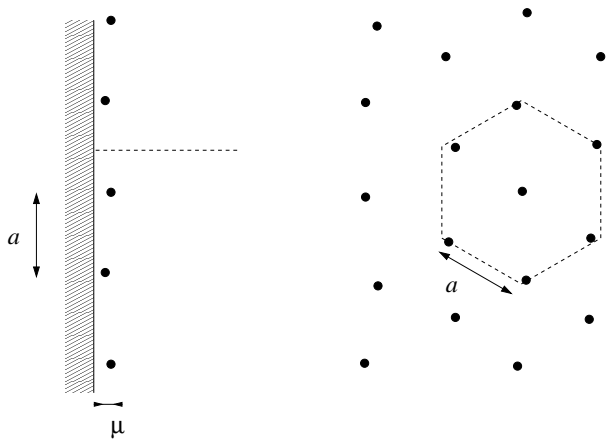}
\caption{(Left) Side view of the ionic atmosphere in the
vicinity of a single charged plate (large $d$ limit). The large value of the
coupling parameter $\Xi$ makes that $\mu \ll a$. (Right) Front view of the
system. When $T=0$ (i.e. $\Xi=\infty$ and the Gouy length $\mu=0$), the counter-ions are exactly located
on the hexagonal lattice with constant $a$. At finite but large $\Xi$
(the situation shown), the
counter-ions are close to these limiting positions. }
\label{fig:oneplate} 
\end{SCfigure}

In the subsequent analysis,
we consider  
$\Xi \gg 1$, where the counter-ions form a quasi-2D layer around the (still
here isolated) plate: indeed, they are then in a configuration close 
to the ground state, which is a hexagonal two dimensional crystal,
the so-called Wigner crystal \cite{Wign34}.
In the ground state ($T=0$ or equivalently $\Xi=\infty$), 
it should be noted that any counter-ion feels
the electric field created by the plate, while the field due to other
counter-ions vanishes by symmetry: 
all ions lie in the same plane. 
Hence, due to thermal excitations and provided $d$ is small
enough, the ions can move slightly away from 
the plate and explore a region of size $\mu$, where the energy cost 
for exciting a given ion, to leading order in $kT$,
stems entirely from the plate potential. Remembering that the
bare plate creates for $z>0$ an electrostatic potential 
$-2\pi \, \sigma e \, z / \epsilon$,
we can therefore 
estimate the value of $\mu$ by writing 
\begin{equation}
q e \, \frac{2 \pi}{\epsilon } \, \sigma e \, \mu \,=\, kT 
\quad \Longrightarrow \quad \mu = \frac{1}{2 \pi q\ \ell_B \sigma} .
\end{equation}
It is instructive to compare this new length scale $\mu$, 
called the Gouy length, to both $a$ and $\ell_B$ \cite{NJMN05}:
from the condition $\Xi \gg 1$, we infer that
\begin{equation}
\mu \ll a \ll q^2 \ell_B \quad \hbox{and more precisely} \quad 
\frac{\mu}{a} \propto \frac{a}{q^2 \ell_B} \propto \frac{1}{\sqrt\Xi}.
\label{eq:ordering}
\end{equation}
The corresponding configuration is shown in 
fig.\ref{fig:oneplate}. We now ascertain the consistency of neglecting
the field due to other counter-ions on a given one:
if the contribution due to other
ions is accounted for, we obtain that an ion moved from its ground state
position experiences an additional potential in $q^2 z^2/(\epsilon a^3)$. For
$z=\mu$, this yields 
$kT \ell_B \, \mu^2/a^3 \propto kT \, \Xi^{-1/2} \ll kT$.
Here as in most of our discussion, the requirement of a large
$\Xi$ is paramount.%
\footnote{These argument immediately provide the ion density profile
around an isolated charged plate, showing that as far as the $z$ coordinate
is concerned and to leading order in $\Xi$, the ions decouple from each other,
thereby forming an ideal gas in an external gravitational-like 
field. The ensuing profile is simply \cite{Shkl99,Netz01,GrNS02,SaTr11PRL}
$\rho(z) \propto \exp(-z/\mu$). This expression should become
asymptotically exact when $\Xi\to\infty$ 
--as validated by numerical simulations-- 
entailing that the profile should obey the contact theorem, 
valid at all couplings. This is indeed the case:
from normalisation, it follows that 
$\rho(z) = 2 \pi \ell_B \sigma^2 \exp(-z/\mu)$ so that
$\rho(0) = 2 \pi \ell_B \sigma^2 $. It should not be forgotten though 
that the ideal gas picture only holds for the $z$ degree of freedom,
whereas the ions strongly repel in the perpendicular direction (parallel
to the plate), where they form a crystal.  
}

Seen from a large distance by a infinitesimal test charge moving along the
$z$ axis as shown by the dashed line in fig.\ref{fig:oneplate}-left,
the whole structure (plate+ions) does not create any electric field
since it is electro-neutral
[more precisely, the electric field is exponentially small for $z \gg a$, and
decays like $\exp(-z/a)$]. However, when the test charge is close to the plate
($z \ll a$), and assuming it does not sit on a counter-ion but 
at maximum possible distance from counter-ions
(see the dashed line in fig.\ref{fig:oneplate}-left), the electric field is that 
of the bare plate, $2 \pi \sigma z/\epsilon$. 
This innocuous remark plays an important role in the argument 
to follow.

\section{From infinite to small inter-plate distances: the unbinding scenario}
\label{sec:smalld}

\begin{SCfigure}[1.8]
\centering
\includegraphics[width=0.17\textwidth]{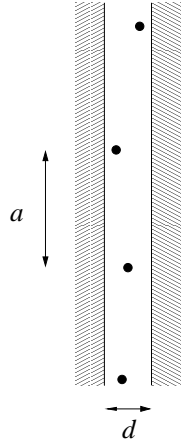}
\caption{The two plates in the small separation limit. 
It is assumed here that
$d \ll a$, which can still be compatible with 
$\mu \ll d$, see eq. (\ref{eq:ordering}).}
\label{fig:geometry_bis} 
\end{SCfigure}

In light of the previous discussion, it is only when $d$ becomes smaller
than the lateral correlation length $a$ that the ionic structure 
on a plate starts to distort compared to the infinite separation case
(this expectation is fully corroborated by a detailed analysis \cite{SaTr12PRB}).
The problem then becomes complex, except when $d \ll a$,
see fig.\ref{fig:geometry_bis}. In this
short separation limit, the ions are confined to a quasi-2D geometry
and we need to identify the position of a given counter-ion 
by its projection parallel to the plates $\bm r_\parallel$,
and its coordinate $z$. Both coordinates
should be considered separately. Projecting the structure
onto a plate, thereby eliminating $z$ and focussing onto the
set of $\bm r_\parallel$ for all ions, we obtain a Wigner crystal
very reminiscent of that sketched in fig.\ref{fig:oneplate}-right,
with the only difference that the surface density of ions 
is now doubled, and that the lattice constant is hence
changed according to $a \to a/\sqrt{2}$.
On the other hand, the behaviour of the $z$ coordinate is more subtle. 

A given counter-ion in the slab 
experiences the sum of the potentials created by the plates
plus a term due to counter-ions interactions.
The first plates-ion contribution yields a vanishing electric field
(constant potential, say 0) in the symmetric plate case.
As for the second ion-ion interaction and
as was the case for the single-plate analysis of section
\ref{sec:larged}, it is a small quantity 
since all counter-ion basically lie in the same plane,
with an ensuing small electric field created.
To be more specific, this inter-ionic energy per particle 
 can be estimated as
being of order $kT \ell_B \,d^2/a^3 \propto kT \,(d/a)^2 \,\Xi^{1/2}$;
it is thus small compared to $kT$ provided that 
$d/a \ll \Xi^{-1/4}$ or equivalently $d/\mu \ll \Xi^{1/4}$.
Under that proviso, we can neglect the ion-ion interactions:
the ions then experience
a uniform potential, freely explore the available space $0 \leq z \leq d$,
and consequently adopt a uniform 
density profile along $z$: $\rho(z)=\hbox{cst}$.
From normalisation $\int \rho \, dz = 2\sigma/q$ and we necessarily have 
$\rho(z) = 2\sigma /(q d)$. There is consequently an {\em unbinding}
of ions which takes place: when $d$ is large, the ions are confined
in a narrow region of extension $\mu$ around each plate,
see section \ref{sec:larged}.
When $d \ll a$, the counter-ions unbind from the plates, 
along the $z$ direction, as a consequence
of the (nearly) cancellation of inter-ionic interactions,
and can explore the full slab width, of extension $d$.
Note that $d$ can be larger than $\mu$, while being small 
compared to $a/\Xi^{1/4}$, see eq. (\ref{eq:ordering}). 
Still, in the transverse direction
$\bm r_\parallel$, the ions are strongly correlated and are
essentially frozen on the hexagonal Wigner positions.

Taking advantage of the unbinding phenomenon, 
the last ingredient of the analysis is to recall the contact 
theorem which allows to cast the inter-plate pressure
in the form
\begin{equation}
\frac{P}{kT} \,=\, \frac{2\sigma}{q d} \, - 
\frac{2 \pi\,\sigma^2 e^2}{\epsilon \,kT}  \, = \, 
2 \pi \ell_B \sigma^2 \left( \frac{2 \mu}{d} -1
\right).
\label{eq:eos}
\end{equation}
Thus, when $d > 2 \mu$, the pressure is negative, and the two like-charge
plates attract. On the other hand, when 
$d<2\mu$, the pressure is repulsive and its positive sign
stems from the penalising entropy for confining ions in too narrow
a slab, which becomes overwhelming and leads to the divergence
$P\sim 2 \sigma kT /(q d)$ when $d \to 0$.
It can be pointed out that a systematic expansion 
in inverse powers of the (large) coupling parameter $\Xi$ 
completely confirms our argument \cite{SaTr11PRL,SaTr11PRE}, 
and provides the following 
correction to the equation of state (\ref{eq:eos})
\begin{equation}
\frac{P}{kT} \,=\, 2 \pi \ell_B \sigma^2 \left( \frac{2 \mu}{d} -1
 +  \frac{d}{a} \,0.6672... + {\cal O}(d^2/a^2)\right)
\label{eq:eossc}
\end{equation}
Here, the same requirement as above holds for $d/a$, namely that
$d/a \ll \Xi^{-1/4}$. The term $d/a$ in the parenthesis 
in eq. (\ref{eq:eossc}) is therefore a small correction 
compared to the others. 
Expression (\ref{eq:eossc}) shows that the ion-ion interactions
that were neglected in deriving eq. (\ref{eq:eos}) contribute 
to a small correction, as expected. In other words,
the true density profile, leading to (\ref{eq:eossc})
via the contact theorem, is not exactly flat as considered,
but slightly peaked in the vicinity of the 
plates\footnote{these two small peaks are precursors of their
ground state counterpart, that develop when $\Xi \gg (a/d)^4$, see the
discussion in section \ref{ssec:ground}.}
\cite{SaTr11PRL,SaTr11PRE}.
In this respect, the equation of state (\ref{eq:eos})
is necessarily a lower bound for the pressure.
It is all the more accurate as the condition $d \ll a/\Xi^{1/4}$
is met, and since we may have $\mu \ll d$, $P$
may take values close to the maximally attractive
bound $-2 \pi \sigma^2 e^2 / \epsilon $. This is a
key point explaining the stability of cement pastes 
\cite{PeCD97}, that can be viewed to first approximation
as made up of parallel charged plates with intercalated
multivalent ions. 

\section{Discussion}
\label{sec:discuss}

We now examine in more details some questions 
raised by our treatment, together with a possible generalisation.

\subsection{Unbinding scenario and ground-state structure}
\label{ssec:ground}
\begin{SCfigure}[1.7]
\centering
\includegraphics[width=0.27\textwidth]{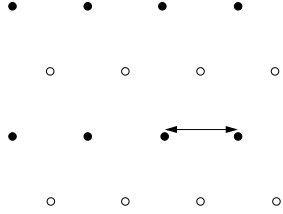}
\caption{Front view of the ground-state structure for $d\ll a$. 
The structure is a hexagonal crystal of lattice constant 
$a/\sqrt{2}$ (shown by the double-arrow), 
where the open and filled symbols correspond to 
counter-ions in contact with the plate at $z=0$ or at $z=d$
respectively.
}
\label{fig:ground} 
\end{SCfigure}

The argument put forward in section \ref{sec:smalld}
for like-charge attraction 
hinges upon the unbinding
of ions from the near vicinity of the plates, which leads
to a uniform $\rho(z)$ profile across the slab.
We have argued that this could occur at large $\Xi$ provided
$d \ll a/\Xi^{1/4}$,
but it is clear that such a flat profile is far from the 
ground state structure of the bilayer. Indeed, Earnshaw theorem
states that an equilibrium position cannot be obtained
in a system of point charges under the action of electrostatic
forces alone \cite{Earn1842}. 
This means here that the counter-ions stick to the
plates when $T=0$, and a more refined 
study reveals that charges adopt the 
pattern shown in fig.\ref{fig:ground} \cite{GoPe96,SaTr12PRB}.
The corresponding $\rho(z)$ profile is thus a sum of two $\delta$
peaks at $z=0$ and $z=d$, very non-uniform\ldots~
There is of course no incompatibility between our
uniform smooth profile and the correct singular ground state.
When the inter-plate distance $d$ and $a$ are fixed (note that both
length scales are temperature independent, unlike the Gouy length), 
and $T$ is decreased
to 0 in order to realize the ground state, $\Xi$ increases
and the condition $d \ll a/\Xi^{1/4}$ is violated at some
point. There is then a crossover between the flat profile and
the strongly peaked one that is reached in the ground state.
In other words, the analysis of section
\ref{sec:smalld} is valid for large $\Xi$, but nevertheless
requires that $\Xi \ll (a/d)^4$. By working with suitably
rescaled quantities, it is nevertheless possible to obtain 
expression that are strictly speaking correct for $\Xi\to\infty$,
see below.

\subsection{Back to the failure of mean-field}
\label{ssec:mean-field}
It is informative to go back to the mean-field treatment,
the Poisson-Boltzmann (PB) theory \cite{Levi02}, which
always predicts repulsion between the plates \cite{Neu99,SaCh00,Triz00}.
Yet, the contact theorem that has been used here, 
is obeyed by Poisson-Boltzmann profiles. Furthermore,
the mean-field profile also becomes flat for small enough $d$.
So, what is going on? The answer simply lies in the fact the
length scale $a$ has no counter-part in the mean-field approach,
where the counter-ions are accounted for via their profiles,
while discarding their discrete nature.
As a consequence, it is only for $d \ll \mu$ that the PB
profile becomes flat \cite{Ande06}. Hence, the equation 
of state (\ref{eq:eos}) also holds for PB theory,
but only when $d \ll \mu$, a regime where $P>0$.

\begin{figure}[htb]
\begin{center}
\centering
\includegraphics[width=0.6\textwidth,clip]{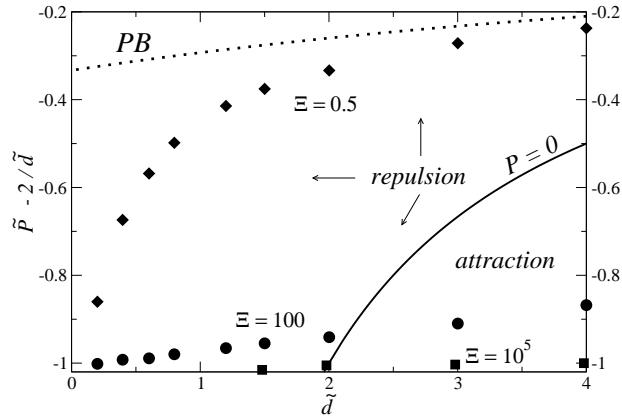}
\caption{Rescaled and shifted 
equation of state as a function of rescaled distance:
$\widetilde P -2/\widetilde d$ is plotted versus $\widetilde d$,
with $\widetilde P = P/(kT 2 \pi \ell_B \sigma^2)$ and $\widetilde d = d/\mu$.
The symbols show the Monte Carlo data of Ref \cite{MoNe02}: the diamonds
are for $\Xi=0.5$ (relatively weak-coupling),
the circles are for $\Xi=100$ (moderately strong-coupling)
and the squares are for $\Xi=10^5$ (very strong-coupling). 
The Poisson-Boltzmann
(mean-field) result 
is shown by the upper dotted line. The continuous curve displays
the locus of points where $P=0$. Hence, below this curve, the like-charge
plates attract.}
\label{fig:Netz16} 
\end{center}
\end{figure}

This leads us to a second remark. It is often stated that
PB theory becomes correct when $\Xi$ is small enough (see e.g. 
\cite{Kenn84} for a proof in a system having both
co- and counter-ions).
Loosely speaking, this is true, but some phenomena 
are governed in the real system (say in a simulation)
by the length scale $a$, and are therefore necessarily missed
at PB level, no matter how small $\Xi$ is. 
This is illustrated in fig.\ref{fig:Netz16},
which shows pressures obtained in Monte Carlo simulations
for three different coupling parameters, $\Xi=0.5$, $\Xi=100$ and
$\Xi=10^5$.
The curve for $\Xi=0.5$ is close to the PB result shown by the dotted 
line provided $d$ is not too small, see also fig.\ref{fig:Netz_eos} below.
For $\Xi=0.5$, we have $a \simeq \mu$, and we see that the
Monte Carlo data depart from the PB prediction for 
$\widetilde d <1$, i.e. $d<a$.
In the rescaled units used in the figure,
our equation of state (\ref{eq:eos}) reads
\begin{equation}
\widetilde P \,=\, \frac{2}{\widetilde d}- 1.
\label{eq:eosrescaled}
\end{equation}
Interestingly, for $d \ll a$ and $\Xi=0.5$ 
(upper symbols) we see that our expression 
is perfectly obeyed, since $\widetilde P -2/\widetilde d \to -1$.
Indeed, the requirement of a flat profile is met for $d \ll \mu$.
On the other hand, the PB result is 
$\widetilde P -2/\widetilde d \to -1/3$ \cite{Netz01}, as 
can be seen in the figure, which significantly 
departs from the simulation data.
It can be concluded that (\ref{eq:eos}) or (\ref{eq:eosrescaled})  are 
not limited to
strongly coupled systems, but also provide the limiting small $d$ behaviour
at arbitrary coupling $\Xi$. 
Yet, we emphasise that (\ref{eq:eos}) or (\ref{eq:eosrescaled}) 
only lead to attractive
behaviour when $\Xi$ is large (attraction requires
having $\mu \ll a$); for $\Xi=100$ or $\Xi=10^5$ and $d>2\mu$ ($\widetilde d >2$)
the pressure does indeed exhibit a negative sign: the circles 
and squares lie below the separatrix $P=0$ shown by the continuous curve 
in the lower right corner of fig.\ref{fig:Netz16}, see also
fig.\ref{fig:Netz_eos}. Moreover, while eq. (\ref{eq:eos})
is fairly well obeyed by the Monte Carlo data at 
$\Xi=100$, the improved version (\ref{eq:eossc})
fares better and is in good agreement with the simulation results,
as discussed in detail in \cite{SaTr11PRE}.
When the electrostatic coupling is increased even further,
the agreement between simulations and eq. (\ref{eq:eos})
becomes excellent, see figures \ref{fig:Netz16} and
\ref{fig:Netz_eos}.

To finish with the scaling form (\ref{eq:eosrescaled}),
we point out that from 
dimensional analysis, the exact rescaled inter-plate pressure 
can in general be written as a function of two arguments only,
i.e. $\widetilde P_{exact}(\widetilde d,\Xi)$. This is true for all 
values of $\Xi$. Working at fixed $\widetilde d$ and letting
$\Xi \to \infty$ to enforce the strong-coupling limit, we obtain 
$d/a \propto \widetilde d \,\Xi^{-1/2}$ which is thus negligible 
compared to $\Xi^{-1/4}$. This means that 
\begin{equation}
\lim_{\Xi\to\infty} \,\widetilde P_{exact}\,(\,\widetilde d\,,\,\Xi\,)
\,=\, \frac{2}{\widetilde d}- 1.
\end{equation}
This is clearly illustrated in fig.\ref{fig:Netz_eos}. 
Of course, for larger distances than those in the figure,
the pressure for $\Xi=10^5$ would start to deviate
from the prediction (\ref{eq:eosrescaled}), since
the criterion $d/a \ll \Xi^{1/4}$ would no longer be met
(in the present case,  $d/a \simeq \Xi^{1/4}$ corresponds
to $\widetilde d \simeq 20$ indicating that deviations
would start to appear for already $\widetilde d $ on the order
of 50 or less). Incidentally, fig.\ref{fig:Netz_eos}
also highlights
a) that PB holds for large enough distances under weak-coupling
(diamonds, $\Xi=0.5$) and 
b) that for large enough $\Xi$, the pressure is negative provided
$\widetilde d > 2$. How large should $\Xi$ be is not provided
by our argument. It has been reported that $\Xi_c \simeq 12$
is the threshold value below which no attraction is possible 
\cite{MoNe02}.

\begin{figure}[htb]
\begin{center}
\centering
\includegraphics[width=0.6\textwidth,clip]{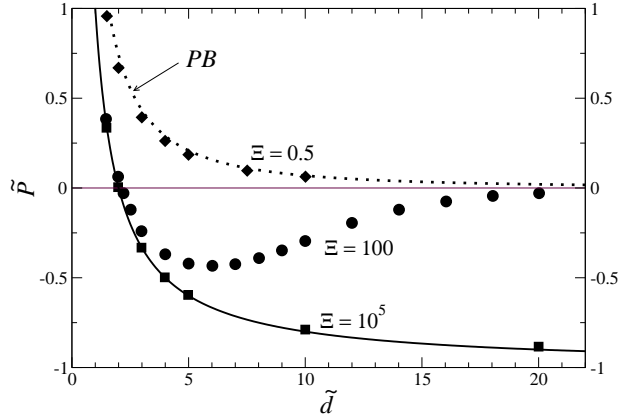}
\caption{Rescaled equation of state $\widetilde P$ versus
$\widetilde d$. The symbols are for the Monte Carlo data 
of Ref \cite{MoNe02} and have the same meaning as in fig.\ref{fig:Netz16}.
They correspond to the numerical evaluation of $P_{exact}$.
The Poisson-Boltzmann
pressure is again shown by the dotted curve and the thick line displays
the equation of state (\ref{eq:eosrescaled}).}
\label{fig:Netz_eos} 
\end{center}
\end{figure}

\subsection{Large distance behaviour}
\label{ssec:larged}
At this point, it is interesting to discuss the phenomenology
for large distances. What is the sign of the pressure? 
We have repeatedly emphasised that our argument is limited
to small enough distances, so that it is not informative for the
large $d$ physics. Worse than that, it appears that there is to date 
no reliable answer --be it experimental, numerical or analytical-- 
to the question raised, and we are confined here to speculation. 
A plausible scenario is that when the two plates are at a large
distance from each other, the ionic profile far enough from the plates
is dilute, so that the local coupling parameter is small
in the interstitial region. Mean-field
can be expected to hold there \cite{Shkl99,ChWe06,DSDL09},
from which we might infer that the large distance pressure should
be positive. Hence, for a given coupling parameter $\Xi$,
we may surmise a reentrant behaviour, with a repulsive pressure at both 
small and large distances, and an attractive window in between.
This expectation may be dimension-dependant,
since a recent two dimensional study of the same problem
has shed doubts on the re-entrance of repulsion at large $d$ \cite{SaTr11EPJE}. 
It should be outlined though that the log-form Coulomb potential takes
in two dimension is scale free, so that the coupling parameter
in 2D is density independent ($\beta q^2$ where $\beta$ is the inverse
temperature, which unlike $\Xi$ does not depend on the macromolecule
charge). This is a notable difference
with 3D systems.

\subsection{Asymmetric plates generalisation}
\label{ssec:assym}
It is straightforward to extend our analysis to the situations
where the two plates bear unequal surface charges. We leave the
derivation as an exercise, emphasising that the main difference
with the treatment put forward in section \ref{sec:smalld}
is that the electric field created by the two plates
is now non-vanishing, but still uniform. It is given by 
$2 \pi (\sigma_1-\sigma_2) e/\epsilon$ where $\sigma_i$ denotes
the surface charge on plate $i$ (plate 1 at $z=0$ and plate 2 
at $z=d$), from which we define the asymmetry
parameter $\zeta = \sigma_2/\sigma_1$. We can again, for
small inter-plate distance neglect the ion-ion interaction.
The ions thus experience a linear potential,
which leads to an exponential profile in $\exp[-z(1-\zeta)/\mu]$,
where $\mu=\mu_1$ is the Gouy length of plate 1. 
Upon proper normalising the counter-ion profile 
($q\int \rho dz = \sigma_1+\sigma_2$), and invoking
once more the contact theorem which reads
\begin{equation}
P \,=\, \rho(0) \, kT \, -\, \frac{2 \pi}{\epsilon} \,\sigma_1^2 \,e^2  = 
\rho(d) \, kT \, -\, \frac{2 \pi}{\epsilon} \,\sigma_2^2 \,e^2 ,
\label{eq:contactgen}
\end{equation}
we obtain \cite{KTNB08,SaTr11PRE,PaTr11}
\begin{equation}
\widetilde P = \frac{ P}{kT 2 \pi \ell_B \sigma_1^2} = -\frac{1}{2}(1+\zeta^2)+
\frac{1}{2}(1-\zeta^2)\coth\left(\frac{(1-\zeta)\,d}{2 \mu}\right) .
\label{eq10}
\end{equation}
This expression reduces to (\ref{eq:eos}) when $\zeta \to 1$, as it should.
An interesting check for the consistency of the approach is to verify
that the same pressure is recovered no matter which plate is used
for evaluating the pressure in eq. (\ref{eq:contactgen}).
Another interesting and exact consequence of the contact theorem
(\ref{eq:contactgen}) is that the interactions between a plate at arbitrary 
charge and a neutral one are always repulsive: $\sigma_2=0$ leads
to $P=\rho(d) \, kT>0$.

\section{Conclusion}
\label{sec:concl}

We considered two interacting parallel charged plates, forming
a slab where neutralising counter-ions are confined. 
We have presented an argument to unveil the mechanisms behind
the like-charge attraction that may ensue at large 
Coulombic coupling parameter $\Xi$, following a mechanical route
whereas other approaches often rely 
on an energy route \cite{RoBl96,Shkl99,Levi02}.
Our argument hinges upon the fact that 
at small enough inter-plate separation, the counter-ions
unbind from the vicinity of the plates. From the contact
theorem, attraction can only set in when the counter-ion density
at contact $\rho(0)$ is below its large distance value,
and the unbinding phenomenon is thereby the key to like-charge
attraction. 
More precisely,
a single counter-ion picture prevails, where due
to strong lateral repulsion (parallel to the plates),
the ions move in the perpendicular $z$ direction under
the potential created by the plates only, 
as if they did not interact. This viewpoint is fruitful
and efficient for the planar geometry considered, but
it is misleading in two respects: first, it may lead to believe
that while the single particle image provides the exact leading
order pressure when $\Xi\to \infty$, the next correction in
the $\Xi$ expansion stems from two-body interactions, in a virial-like
scheme. This is the essence of the virial approach of Refs.
\cite{Netz01,MoNe02,NJMN05}, which has been shown to be incorrect 
\cite{SaTr11PRL,SaTr11PRE}. Second, the single particle picture
is in general inappropriate, even to obtain the dominant behaviour
at large $\Xi$. This appears most clearly when considering
the ionic profile for a
single plate defining two half spaces with different dielectric
constants (work in preparation).

To dominant order in the coupling parameter $\Xi$, the fact that
there is an underlying Wigner crystal of ions has no apparent signature
on the interactions
(this signature should be sought in the next to dominant term 
\cite{SaTr11PRL,SaTr11PRE}). However, the presence of the crystal
is important for allowing the unbinding of ions and in this respect,
the description of strongly coupled charged matter is in the realm
of low temperature physics, as discussed in the review \cite{GrNS02}. 
It is then instructive to remind the value of the coupling parameter 
where crystallisation occurs, which is on the order of
$\Xi_c \simeq 3.10^5$ \cite{BaHa80}. If strong coupling approaches 
of the Wigner kind, such as that leading to (\ref{eq:eossc})
where restricted to this range of coupling, they would be
of little interest, since it is in practice
difficult to exceed $\Xi=100$ (e.g. trivalent counter-ions
are required in water for already large surface charges
such that $\sigma \ell_B^2$ is of order 1 \cite{NJMN05}). 
However, it has been shown that the
predictions derived from the Wigner picture apply ``down to''
much smaller value than $\Xi_c$, such as 50 or sometimes less
\cite{SaTr11PRE}. This means that the key point is not
the crystal in itself, but the existence of a strong
correlation hole around each counter-ion.

Acknowledgements: The authors would like to acknowledge interesting
conversations with Y. Levin, M. Kandu\v{c}, J.P. Mallarino, A. Naji, R. Netz
R. Podgornik and G. T\'ellez.

\bibliographystyle{varenna}
\bibliography{bibliog}

\end{document}